\documentclass[runningheads]{llncs}
\usepackage{graphicx}
\usepackage{cite} 
\usepackage{times}
\usepackage{epsfig}
\usepackage{amsmath}
\usepackage{amssymb}
\usepackage{mwe}
\usepackage{acro}
\usepackage{amssymb}
\usepackage{xcolor,colortbl}
\usepackage{tabularx}
\usepackage{relsize}
\usepackage{pifont}
\usepackage{booktabs} 
\usepackage{multirow}
\usepackage{multicol}
\usepackage{adjustbox}
\usepackage{float}
\usepackage[colorlinks,
            linkcolor=red,  
            anchorcolor=blue, 
            citecolor=green,   
            ]{hyperref}
\usepackage[flushleft]{threeparttable}

\begin{document}
\title{Democratizing Pathological Image Segmentation with Lay Annotators via Molecular-empowered Learning}

\author{Ruining Deng\inst{1} \and
Yanwei Li\inst{1} \and
Peize Li\inst{1} \and
Jiacheng Wang \inst{1}\and
Lucas W. Remedios \inst{1}\and 
Saydolimkhon Agzamkhodjaev \inst{1} \and
Zuhayr Asad \inst{1} \and
Quan Liu \inst{1} \and
Can Cui\inst{1} \and
Yaohong Wang \inst{3} \and
Yihan Wang \inst{3} \and 
Yucheng Tang \inst{2} \and
Haichun Yang \inst{3} \and
Yuankai Huo \inst{1}}

% index{Deng, Ruining}
% index{Li, Yanwei}
% index{Li, Peize}
% index{Wang, Jiacheng}
% index{Remedios, Lucas}
% index{Agzamkhodjaev, Saydolimkhon}
% index{Asad, Zuhayr}
% index{Liu, Quan}
% index{Cui, Can}
% index{Wang, Yaohong}
% index{Wang, Yihan}
% index{Tang, Yucheng}
% index{Yang, Haichun}
% index{Huo, Yuankai}

% \authorrunning{H. Yang et al.}
% % 1107
% \author{submission 702}
% \institute{******}

\institute{
1. Vanderbilt University, Nashville TN 37215, USA, \\
2. NVIDIA Corporation, Santa Clara and Bethesda, USA\\
3. Vanderbilt University Medical Center, Nashville TN 37232, USA, \\
}

%email{yuankai.huo@vanderbilt.edu}

\maketitle              % typeset the header of the contribution
\begin{abstract}
Multi-class cell segmentation in high-resolution Giga-pixel whole slide images (WSI) is critical for various clinical applications. Training such an AI model typically requires labor-intensive pixel-wise manual annotation from experienced domain experts (e.g., pathologists). Moreover, such annotation is error-prone when differentiating fine-grained cell types (e.g., podocyte and mesangial cells) via the naked human eye. In this study, we assess the feasibility of democratizing pathological AI deployment by only using lay annotators (annotators without medical domain knowledge). The contribution of this paper is threefold: (1) We proposed a molecular-empowered learning scheme for multi-class cell segmentation using partial labels from lay annotators; (2) The proposed method integrated Giga-pixel level molecular-morphology cross-modality registration, molecular-informed annotation, and molecular-oriented segmentation model, so as to achieve significantly superior performance via 3 lay annotators as compared with 2 experienced pathologists; (3) A deep corrective learning (learning with imperfect label) method is proposed to further improve the segmentation performance using partially annotated noisy data. From the experimental results, our learning method achieved F1 = 0.8496 using molecular-informed annotations from lay annotators, which is better than conventional morphology-based annotations (F1 = 0.7015) from experienced pathologists. Our method democratizes the development of a pathological segmentation deep model to the lay annotator level, which consequently scales up the learning process similar to a non-medical computer vision task. The official implementation and cell annotations are publicly available at \url{https://github.com/hrlblab/MolecularEL}.

\keywords{Image annotation \and Registration \and Noisy label learning \and Pathology.}
\end{abstract}

\section{Introduction}

\begin{figure*}[t]
\begin{center}
\includegraphics[width=0.8\textwidth]{{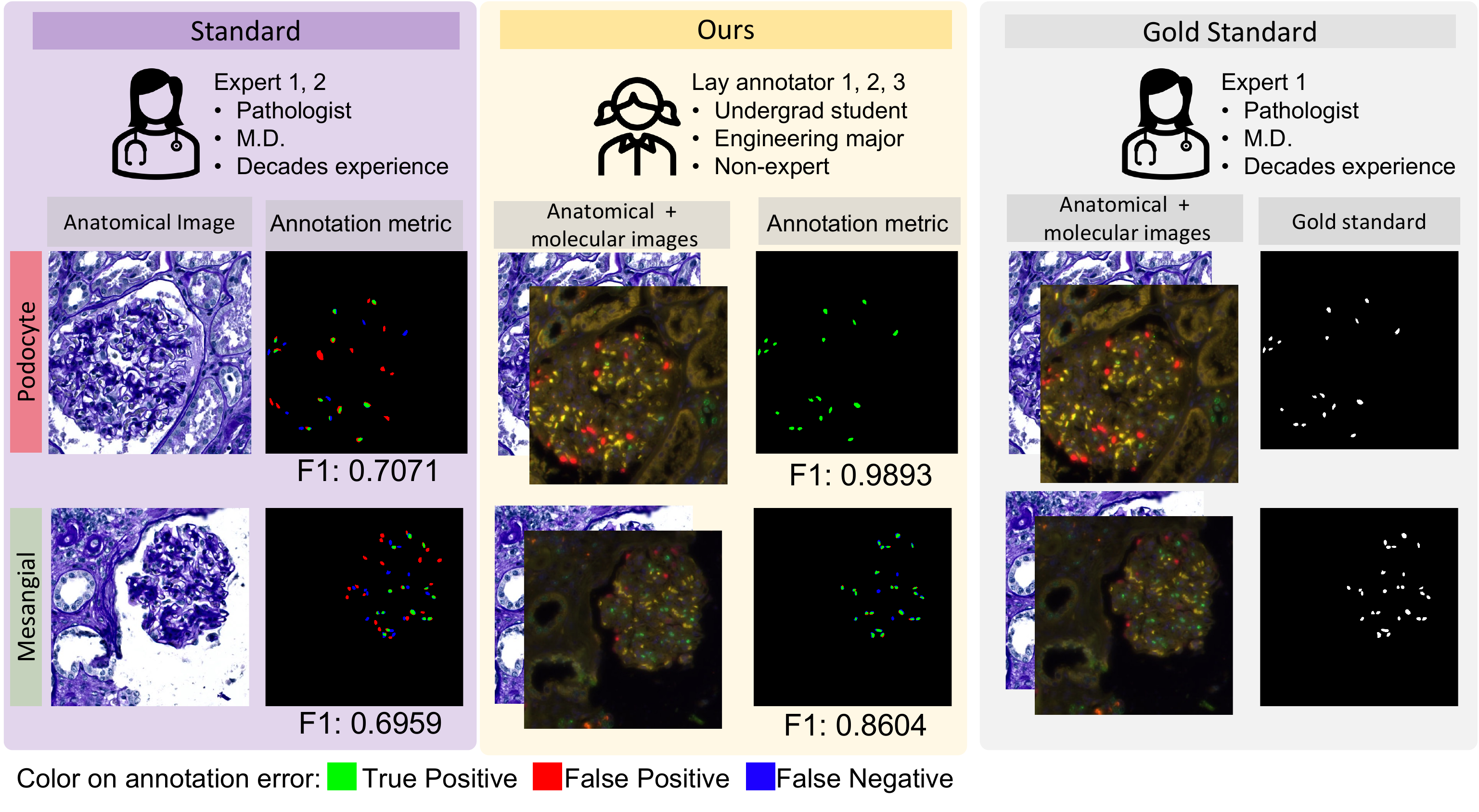}}
\end{center}
\caption{\textbf{The overall idea of this work.} The left panel shows the standard annotation process (PAS only) for developing pathological segmentation models. The middle panel shows our molecular-informed annotation (with both PAS and IF images) that allows for better annotation quality from lay annotators as compared with the left panel. The right panel presents the gold standard annotation for this study, where the annotations are obtained by experienced pathologists upon both PAS and IF images.} 
\label{fig1:Problem}
\end{figure*}

Multi-class cell segmentation is an essential technique for analyzing tissue samples in digital pathology. Accurate cell quantification assists pathologists in identifying and diagnosing diseases~\cite{comaniciu2002cell,xing2016robust} as well as obtaining detailed information about the progression of the disease~\cite{olindo2005htlv}, its severity~\cite{wijeratne2018quantification}, and the effectiveness of treatment~\cite{jimenez2006mast}. For example, the distribution and density of podocyte and mesangial cells in the glomerulus offer a faint signal of functional injury in renal pathology~\cite{imig2022interactions}. The cell-level characterization is challenging for experienced pathologists due to the decades of expensive medical training, long annotation time, large variability~\cite{zheng2021deep}, and low accuracy, while it is impractical to hire massive experienced pathologists for cell annotation. 

% Previously, pathologists had to manually quantify all of the cells according to morphology on pathological images, leading to two primary challenges for cell-level quantification: 1) The workload of labeling the high-resolution giga-pixel pathological image is substantial and high-threshold, requiring around 7 hours (even with decades of clinical experience and professional knowledge) for each tissue sample; (2) Meanwhile, morphology and the spatial relationship among various cells on pathological images are difficult to differentiate, leading to large variability~\cite{zheng2021deep} and low accuracy (in Fig.~\ref{fig1:Problem}).

%~\cite{korzynska2021review,janowczyk2019histoqc,chen2020computer,das2020computer}

Previous works proposed several computer vision tools to perform automated or semi-automated cell segmentation on pathological images~\cite{korzynska2021review}, including AnnotatorJ~\cite{hollandi2020annotatorj}, NuClick~\cite{koohbanani2020nuclick}, QuPath~\cite{bankhead2017qupath}, etc. Such software is able to mark nuclei, cells, and multi-cellular structures by compiling pre-trained segmentation models~\cite{graham2019hover}, color deconvolution~\cite{ruifrok2001quantification}, or statistical analysis~\cite{oberg2012statistical}. However, those automatic approaches still heavily rely on the morphology of cells from pathological Periodic acid–Schiff (PAS) images, thus demanding intensive human intervention for extra supervision and correction. Recently, immunofluorescence (IF) staining imaging has been widely used to visualize multiple biomolecules simultaneously in a single sample using fluorescently labeled antibodies~\cite{day2013fluorescently,moore2017effects}. Such technology can accurately serve as a guide to studying the heterogeneity of cellular populations, providing reliable information for cell annotation. Furthermore, crowd-sourcing technologies~\cite{hsueh2009data,marzahl2020crowd,amgad2022nucls} were introduced generate better annotation for AI learning from multiple annotations.

% However, there are limited - if any - existing approaches that assess the feasibility of developing pathological AI models via lay annotators by infusing molecular images with pathological images.

% Moreover, even with guidance from molecular images, the annotation results from the lay annotators might still be unreliable and error-prone due to the absence of domain expertise and the variability in the quality of staining in molecular images. Therefore, training AI models using such imperfect datasets might result in biased or inaccurate predictions. Several prior arts have tried to reduce the impact of noisy labels, employing strategies such as using multiple annotators and taking a consensus of their labels~\cite{zhan2019learning,long2015multi} or using efficient learning strategies such as confidence score~\cite{northcutt2021confident} or partial label loss~\cite{fan2022pointly}.

In this paper, we proposed a holistic molecular-empowered learning scheme that democratizes AI pathological image segmentation by employing only lay annotators (Fig.~\ref{fig1:Problem}). The learning pipeline consists of (1) morphology-molecular multi-modality image registration, (2) molecular-informed layman annotation, and (3) molecular-oriented corrective learning. The pipeline alleviates the difficulties at the R$\&$D from the expert level (e.g., experienced pathologists) while relegating annotation to the lay annotator level (e.g., non-expert undergraduate students), all while enhancing both the accuracy and efficiency of the cell-level annotations. An efficient semi-supervised learning strategy is proposed to offset the impact of noisy label learning on lay annotations. The contribution of this paper is three-fold:

$\bullet$ We propose a molecular-empowered learning scheme for multi-class cell segmentation using partial labels from lay annotators;

$\bullet$ The molecular-empowered learning scheme integrates (1) Giga-pixel level molecular-morphology cross-modality registration, (2) molecular-informed annotation, and (3) molecular-oriented segmentation model to achieve statistically a significantly superior performance via lay annotators as compared with experienced pathologists;

$\bullet$ A deep corrective learning method is proposed to further maximize the cell segmentation accuracy using partially annotated noisy annotation from lay annotators.

% The official implementation and cell annotation are publicly available at \url{https://github.com/hrlblab/Cell-Seg}.

% \begin{figure}[t]
% \begin{center}
% \includegraphics[width=1\textwidth]{{fig1new.pdf}}
% \end{center}
% \caption{\textbf{Model pipeline.} In the top section, self-supervised model is pretrained with pathology WSIs and finetuned on pathology images for survival prediction. Middle section is supervised pretrained model with natural images of ImageNet and finetuned on pathology images. Our JRT method aggregates both pretrained models to achieve better downstream task performance.} 
% \label{fig:model}
% \end{figure}

\section{Methods}
\begin{figure}[t]
\begin{center}
\includegraphics[width=0.8\textwidth]{{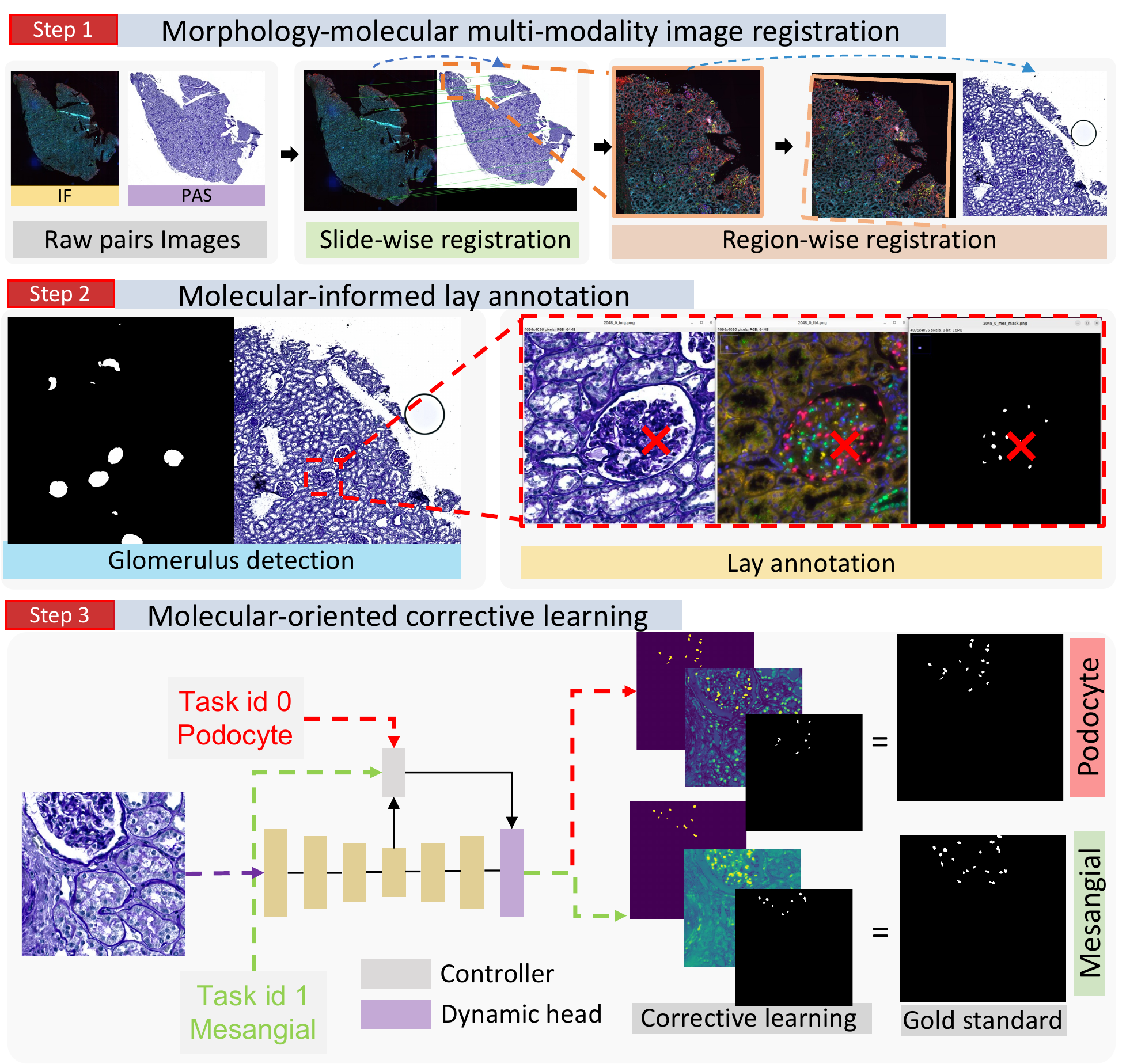}}
\end{center}
\caption{\textbf{The framework of the proposed molecular-empowered learning scheme.} The molecular-empowered learning pipeline consists of (1) morphology-molecular multi-modality image registration, (2) molecular-informed layman annotation, and (3) molecular-oriented corrective learning. It democratizes AI pathological image segmentation by employing only lay annotators.}
\label{fig2:Pipeline}
\end{figure}

The overall pipeline of the entire labeling and auto-quantification pipeline is presented in Fig.~\ref{fig2:Pipeline}. Molecular images are aligned with anatomical images in order to provide accurate guidance for cell labeling by using multi-scale registration. After this registration, a functional unit segmentation model is implemented to localize the regions of glomeruli. Within those glomeruli, lay annotators label multiple cell types by using the pair-wise molecular images and anatomical images in ImageJ~\cite{hollandi2020annotatorj}. A partial-label learning model with a molecular-oriented corrective learning strategy is employed so as to diminish the gap between labels from lay annotators and gold standard labels. 

\subsection{Morphology-molecular multi-modality registration}

Multi-modality, multi-scale registration is deployed to ensure the pixel-to-pixel correspondence (alignment) between molecular IF and PAS images at both the WSI and regional levels. To maintain the morphological characteristics of the functional unit structure, a slide-wise multi-modality registration pipeline (Map3D)~\cite{deng2021map3d} is employed to register the molecular images to anatomical images. The first stage is global alignment. The Map3D approach was employed to achieve reliable translation on WSIs when encountering missing tissues and staining variations. The output of this stage is a pair-wise affine matrix $M_{Map3D}(t)$ from Eq.~\eqref{Map3D}.

\begin{equation}
M_{Map3D} = \arg \min \sum_{i=1}^N||A(x_{i}^{IF},M)- x_{i}^{PAS}||_{Aff_{Map3D}}
\label{Map3D}
\end{equation}

To achieve a more precise pixel-level correspondence, Autograd Image Registration Laboratory (AIRLab)~\cite{sandkuhler2018airlab} was utilized to calibrate the registration performance at the second stage. The output of this step is $M_{AIRLab}(t)$ from Eq.~\eqref{AIRLab}.

\begin{equation}
M_{AIRLab} = \arg \min A_{M_{Map3D}} \sum_{i=1}^N ||A(x_{i}^{IF},M)-x_{i}^{PAS}||_{Aff_{AIRLab}}
\label{AIRLab}
\end{equation}

\noindent where $i$ is the index of pixel $x_i$ in the image $I$, with $N$ pixels. The two-stage registration (Map3D + AIRLab) affine matrix for each pair is presented in Eq.~\eqref{affine}.

\begin{equation}
M = (M_{Map3D},M_{AIRLab})
\label{affine}
\end{equation}

In Eq.~\eqref{Map3D} and ~\eqref{AIRLab}, $A$ indicates the affine registration. The affine matrix $M_{Map3D}(t)$ from Map3D is applied to obtain pair-wise image regions. The $||.||_{Aff_Map3D}$ and \\ $||.||_{Aff_AIRLab}$ in Eq.~\eqref{Map3D} and ~\eqref{AIRLab} indicates the different similarity metrics for two affine registrations, respectively. 

\subsection{Molecular-informed annotation}
After aligning molecular images with PAS images, an automatic multi-class functional units segmentation pipeline Omni-Seg~\cite{deng2022single} is deployed to locate the tuft unit on the images. With the tuft masks, the molecular images then manifest heterogeneous cells with different color signals on pathological images during the molecular-informed annotation. Each anatomical image attains a binary mask for each cell type, in the form of a partial label. Following the same process, the pathologist examines both anatomical images and molecular images to generate a gold standard for this study (Fig.~\ref{fig1:Problem}).

%Specifically, the lay annotator plots the boundaries of the cells on the pathological images with red stains in the molecular image as podocyte cells, and the cells with green stains as mesangial cells.

\subsection{Molecular-oriented corrective learning for partial label segmentation}
\begin{figure}[t]
\begin{center}
\includegraphics[width=1.0\textwidth]{{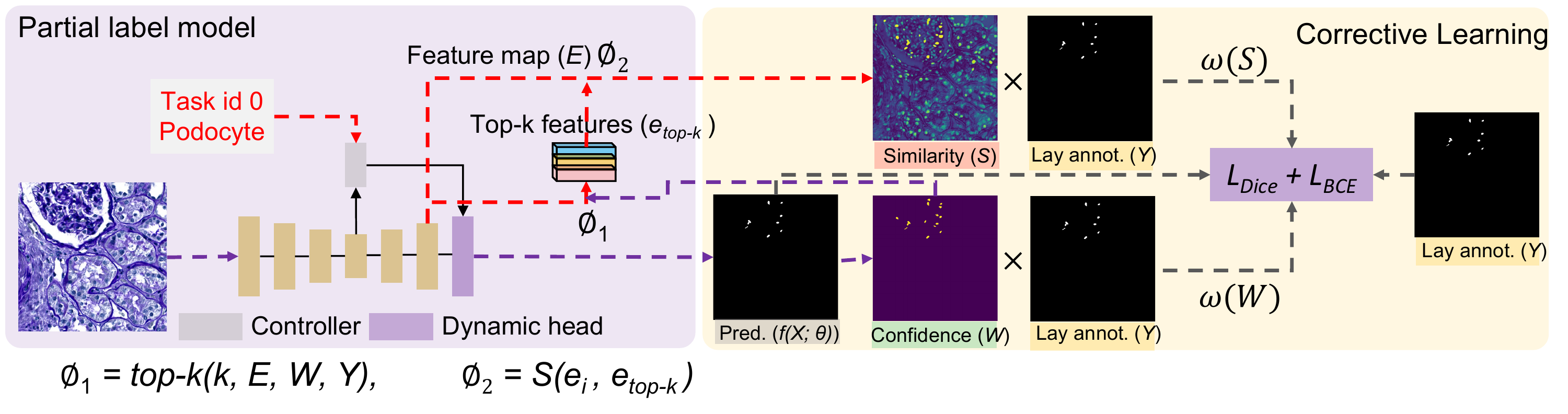}}
\end{center}
\caption{\textbf{The molecuar-oriented corrective learning in partial label model.} A corrective learning are applied to highlight the regions where both the model and lay annotation agree on the current cell type, when calculating the loss function.}
\label{fig3:Corrective}
\end{figure}

The lack of molecular expertise as well as the variability in the quality of staining in molecular images can cause annotations provided by non-specialists to be unreliable and error-prone. Therefore, we propose a corrective learning strategy (in Fig.~\ref{fig3:Corrective}) to efficiently train the model with noise labels, so as to achieve the comparable performance of training the same model using the gold standard annotations.

%Consequently, the utilization of imperfect datasets may result in biased or inaccurate predictions for training a deep learning model. 

%The hypothesis for molecular corrective learning is that the latent features share homogeneity on the pixels of the same cell~\cite{northcutt2021confident}. This similarity can empower the distillation of correct features for each cell.

 Inspired by confidence learning~\cite{northcutt2021confident} and similarity attention~\cite{li2021dual}, top-k pixel feature embeddings at the annotation regions with higher confidences from the prediction probability ($W$, defined as confidence score in Eq.~\eqref{preds}) are selected as critical representations for the current cell type from the decoder(in Eq.~\eqref{topk}). 

\begin{equation}
W = f(X; \theta)[:,1] 
\label{preds}
\end{equation}
\begin{equation}
top-k(k, E, W, Y) = {(e_1, w_1), (e_2, w_2), ..., (e_k, w_k)} \cap Y \in (E, W)
\label{topk}
\end{equation}

\noindent where $k$ denotes the number of selected embedding features. $E$ is the embedding map from the last layer of the decoder, while $Y$ is the lay annotation.

We then implement a cosine similarity score $S$ between the embedding from an arbitrary pixel to those from critical embedding features as Eq.~\eqref{sim}.

\begin{equation}
S(e_i, e_{top-k}) = \frac{\sum_{m=1}^{M}(e_i \times e_{top-k})}{\sqrt{\sum_{m=1}^{M}(e_i)^2} \times \sqrt{\sum_{m=1}^{M}(e_{top-k})^2}}
\label{sim}
\end{equation}

\noindent where $m$ denotes the channel of the feature embeddings.

Since the labels from lay annotators might be noisy and erroneous, the $W$ and $S$ are applied in following Eq.~\eqref{weight_W_S} to highlight the regions where both the model and lay annotation agree on the current cell type, when calculating the loss function in Eq.~\eqref{loss}.

\begin{equation}
\omega(W) = \exp(W) \times Y, \, \omega(S) = S \times Y
\label{weight_W_S}
\end{equation}

\begin{equation}
\mathcal{L}(Y, f(X;\theta)) = (\mathcal{L}_{Dice}(Y, f(X;\theta))) + \mathcal{L}_{BCE}(Y, f(X;\theta)))) \times \omega(W) \times \omega(S)
\label{loss}
\end{equation}

\section{Data and Experiments}
\textbf{Data.} 11 PAS staining WSIs, including 3 injured glomerulus slides, were collected with pair-wise IF images for the process. The stained tissues were scanned at a 20$\times$ magnification. After multi-modality multi-scale registration, 1,147 patches for podocyte cells, and 789 patches for mesangial cells were generated and annotated. Each patch has 512$\times$512 pixels.

%45 4096$\times$4096 pixels image regions were extracted from WSIs for further labeling and model training. Those image regions then generated 1,147 patches for podocyte cells, and 789 patches for mesangial cells. Each patch has 512$\times$512 pixels.

\textbf{Morphology-molecular multi-modality registration.} The slide-level global translation from Map3D was deployed at a 5$\times$ magnification, which is 2 $\mu$m per pixel. The 4096$\times$4096 pixels PAS image regions with 1024 pixels overlapping were tiled on anatomical WSIs at a 20$\times$ magnification, which is 0.5 $\mu$m per pixel.

\textbf{Molecular-empowered annotation.} The automatic tuft segmentation and molecular knowledge images assisted the lay annotators with identifying glomeruli and cells. ImageJ (version v1.53t) was used throughout the entire annotation process. ``Synchronize Windows" was used to display cursors across the modalities with spatial correlations for annotation. ``ROI Manager" was used to store all of the cell binary masks for each cell type.

\textbf{Molecular-oriented corrective learning.}
Patches were randomly split into training, validation, and testing sets - with a ratio of 6:1:3, respectively - at the WSI level. The distribution of injured glomeruli and normal glomeruli were balanced in the split.

\textbf{Experimental setting.} 2 experienced pathologists and 3 lay annotators without any specialized knowledge were included in the experiment. All anatomical and molecular patches of glomerular structures are extracted from WSI on a workstation equipped with a 12-core Intel Xeon W-2265 Processor, and NVIDIA RTXA6000 GPU. An 8-core AMD Ryzen 7 5800X Processor workstation with XP-PEN Artist 15.6 Pro Wacom is used for drawing the contour of each cell. Annotating 1 cell type on 1 WSI requires 9 hours, while staining and scanning 24 IF WSIs (as a batch) requires 3 hours. The experimental setup for the 2 experts and the 3 lay annotators is kept strictly the same to ensure a fair comparison. 

% Stochastic Gradient Descent (SGD) was used for weight update with a learning rate of 0.001 and 0.99 decay. 

\textbf{Evaluation metrics.} 100 patches from the testing set with a balanced number of injuries and normal glomeruli were captured by the pathologists for evaluating morphology-based annotation and molecuar-informed annotation. The annotation from one pathologist (over 20 years’ experience) with both anatomical and molecular images as gold standard (Fig.~\ref{fig1:Problem}). The balanced F-score (F1) was used as the major metric for this study. The Fleiss' kappa was used to compute the inter-rater variability between experts and lay annotators.

\begin{figure*}[t]
\begin{center}
\includegraphics[width=0.8\textwidth]{{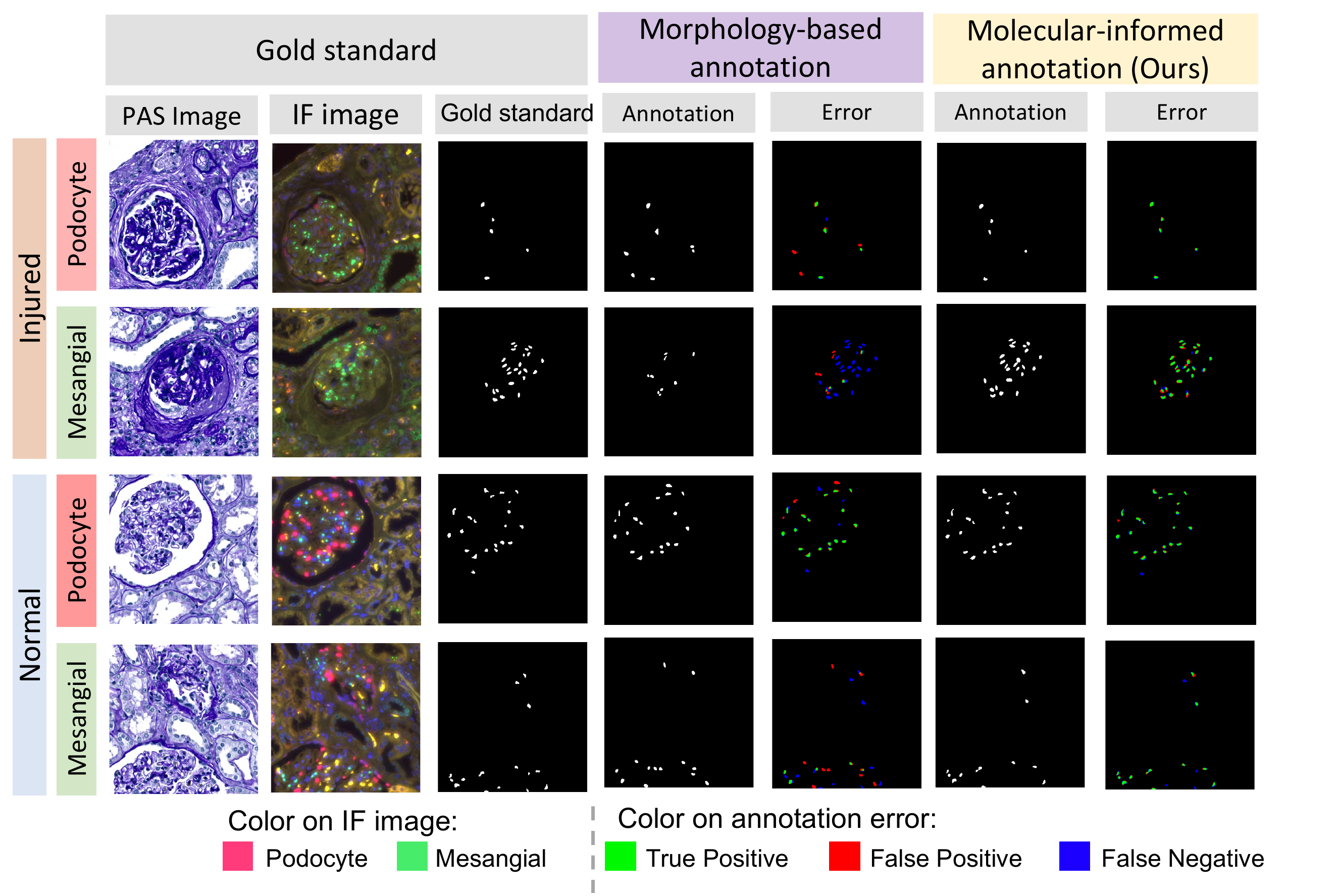}}
\end{center}
\caption{\textbf{Annotation accuracy using learning different strategies.} This figure compares the annotation performance using different strategies. Note that the molecular-informed annotation only employed lay annotators, while the remaining results were from an experienced renal pathologist. } 
\label{fig3:Annotation}
\end{figure*}

\begin{figure*}[t]
\begin{center}
\includegraphics[width=0.5\textwidth]{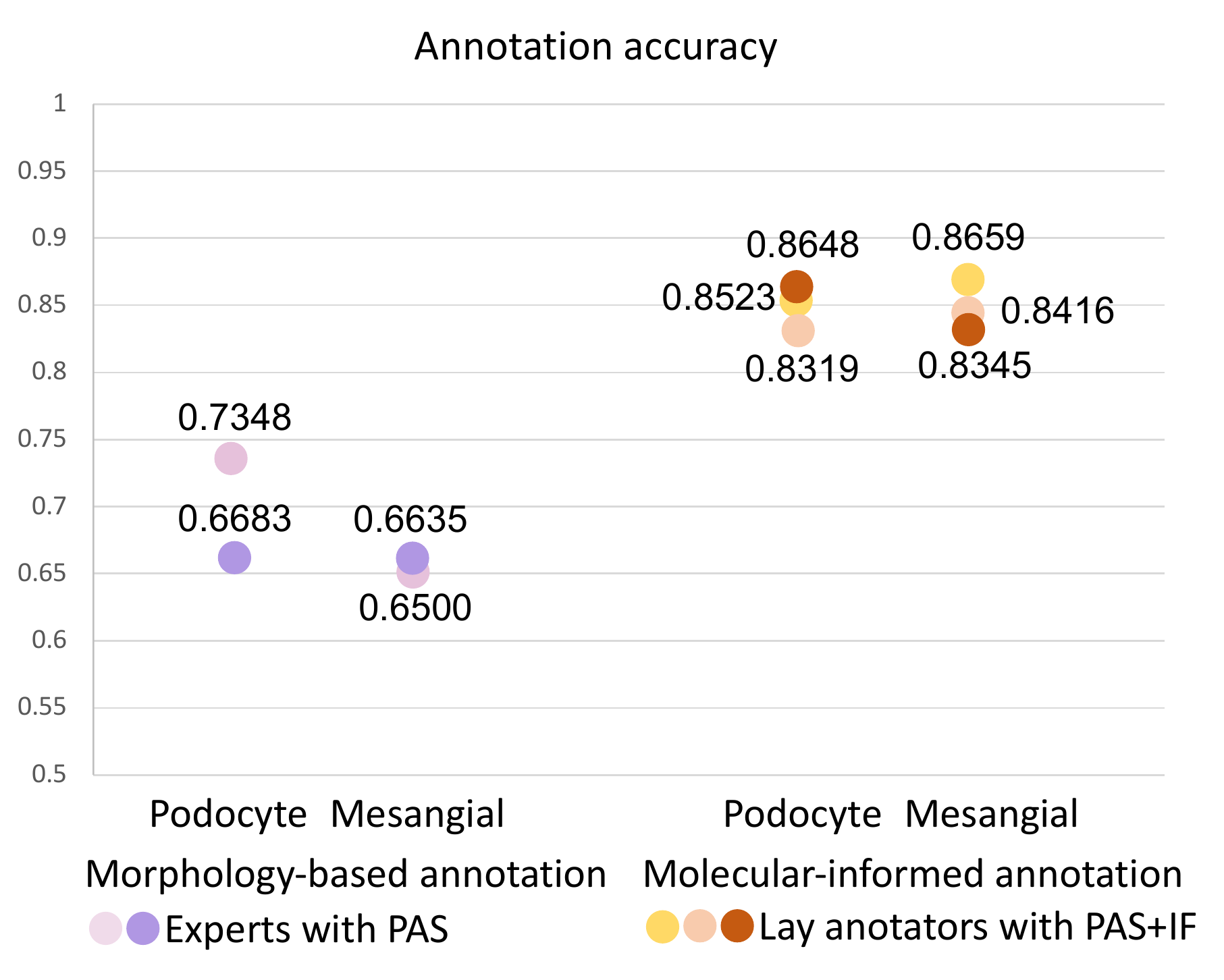}
\end{center}
\caption{\textbf{Annotation accuracy between 2 experts and 3 lay annotators.} This figure compares the annotation performance between morphology-based annotation by 2 experts and molecular-informed annotation by 3 lay annotators. Overall, the molecular-informed annotation achieved better F1 scores than morphology-based annotation.}
\label{fig4:inter-rater}
\end{figure*}

%The models with the best performances in testing were selected within 100 epochs. All the experiments were completed at the same workstation, with the NVIDIA RTX A6000 GPU.

\begin{table*}[t]
% \begin{tabular}{llllllll}
\centering
\caption{
Annotation accuracy from only anatomical morphology and molecular-informed annotation. Average F1 scores and Fleiss' kappa between 2 experts and 3 lay annotators are reported.}
\begin{adjustbox}{width=0.95\textwidth}
\begin{tabular}{l|cccccccc}
\toprule
\multirow{2}{0.8in}{Method} & \multicolumn{2}{c}{Injured glomeruli} & \multicolumn{2}{c}{Normal glomeruli} & \multicolumn{2}{c}{Average} & \multicolumn{2}{c}{Fleiss' kappa}\\
\cmidrule(lr){2-3}
\cmidrule(lr){4-5}
\cmidrule(lr){6-7}
\cmidrule(lr){8-9}
 & Podocyte & Mesangial & Podocyte & Mesangial  & Podocyte & Mesangial & Podocyte & Mesangial \\
\midrule
Morphology-based annotation \\ (2 pathologists with PAS) & 0.6964 & 0.6941 & 0.7067 & 0.6208 & 0.7015 & 0.6567 & 0.3973 & 0.4161  \\
\midrule
Molecular-informed annotation\\ (3 lay annotators with PAS+IF)& \textbf{0.8374} & \textbf{0.8434} & \textbf{0.8619} & \textbf{0.8511} & \textbf{0.8496} & \textbf{0.8473} & \textbf{0.6406} & \textbf{0.5978} \\
\midrule
$p$-value & p$<$0.001 & p$<$0.001& p$<$0.001 & p$<$0.001 &  p$<$0.001&p$<$0.001 & N/A & N/A\\
\bottomrule

\end{tabular}
\end{adjustbox}
\label{tab:CellAnnotation}
\end{table*}

\begin{table*}[h]
% \begin{tabular}{llllllll}
%\centering
\begin{center}
\caption{
Performance of deep learning based multi-class cell segmentation. F1 are reported.  }
\begin{adjustbox}{width=0.8\textwidth}
\begin{tabular}{ll|cccccccccccc}
\toprule
\multirow{2}{0.8in}{Method} & \multirow{2}{0.3in}{Data} & \multicolumn{2}{c}{Injured glomeruli} & \multicolumn{2}{c}{Normal glomeruli} & \multicolumn{2}{c}{Average}\\
\cmidrule(lr){3-4}
\cmidrule(lr){5-6}
\cmidrule(lr){7-8}
 & & Podocyte & Mesangial & Podocyte & Mesangial  & Podocyte & Mesangial \\
\midrule
U-Nets~\cite{ronneberger2015u} & G.S. & 0.6719 & 0.6867 & 0.7203 & 0.6229 & 0.6944 & 0.6617 \\
DeepLabV3s~\cite{chen2017rethinking}& G.S. & \textbf{0.7127} & 0.6680 & 0.7395 & 0.6163 & 0.7251 & 0.6476\\
Residual U-Nets~\cite{salvi2021automated}& G.S. & 0.6968 & 0.6913 & 0.7481 & 0.6601 & 0.7207 & 0.6790\\
\midrule
Multi-class~\cite{gonzalez2018multi}& G.S. & 0.5201 &0.4984 & 0.4992 & 0.4993& 0.5214 & 0.4987 \\
Multi-kidney~\cite{bouteldja2021deep}& G.S.& 0.6735 &0.6734 &0.7542 & 0.6581 & 0.7108 & 0.6691 \\
Omni-Seg~\cite{deng2022single} & G.S. & 0.7115 &  \textbf{0.6970} &  \textbf{0.7746} &  \textbf{0.6895} &  \textbf{0.7407} &  \textbf{0.6940} \\ 
\midrule
Omni-Seg~\cite{deng2022single} & L.A. & 0.6941 & 0.7083 &  \textbf{0.7703} & 0.6822 & 0.7295 & 0.6980 \\
CL~\cite{northcutt2021confident} & L.A. & 0.7047 & 0.6961 & 0.7536 & 0.6754 & 0.7274 & 0.6879 \\
PLL~\cite{fan2022pointly} & L.A. & 0.6276 & 0.6853 & 0.6825 & 0.6268 & 0.6531 & 0.6622\\
MOCL(Ours) & L.A. &  \textbf{0.7198} &  \textbf{0.7157} & 0.7657 &  \textbf{0.6830} &  \textbf{0.7411} &  \textbf{0.7028}\\
\bottomrule
\end{tabular}
\end{adjustbox}
\end{center}
\text{*G.S. denotes gold standard dataset, *L.A. denotes lay annotation dataset}\\
\label{tab:Noisydataset}
\end{table*}

\section{Results}
% \subsection{Accuracy on multi-class cell annotation}

Fig.~\ref{fig3:Annotation}, Fig.~\ref{fig4:inter-rater} and Table~\ref{tab:CellAnnotation} indicate the annotation performance from the naked human eye with expert knowledge and the lay annotator with molecular-informed learning. As shown, our learning method achieved better annotation with higher F1 scores with fewer false positive and false negative regions as compared with the pathologist's annotations. Statistically, the Fleiss' kappa test shows that the molecular-informed annotation by lay-annotators has higher annotation agreements than the morphology-based annotation by experts. This demonstrates the benefits of reducing the expertise requirement to a layman's level and improving accuracy in pathological cell annotation.

\subsection{Performance on multi-class cell segmentation}
In Table.2, we compared the proposed partial label segmentation method to baseline models, including (1) multiple individual models (U-Nets~\cite{ronneberger2015u}, DeepLabv3s~\cite{chen2017rethinking}, and Residual U-Nets~\cite{salvi2021automated}), (2) multi-head models (Multi-class~\cite{gonzalez2018multi}, Multi-Kidney~\cite{bouteldja2021deep}), and (3) single dynamic networks with noisy label learning (Omni-Seg~\cite{deng2022single}). Our results found that the partial label paradigm shows superior performance on multi-class cell segmentation. The proposed model particularly demonstrates better quantification in the normal glomeruli, which contain large amounts of cells.

To evaluate the performance of molecular-oriented corrective learning on imperfect lay annotation, we also implemented two noisy label learning strategies Confidence Learning (CL)~\cite{northcutt2021confident} and Partial Label Loss (PLL)~\cite{li2021dual} with the proposed Molecular-oriented corrective learning (MOCL) on our proposed partial label model. As a result, the proposed molecular-oriented corrective learning alleviated the error between lay annotation and the gold standard in the learning stage, especially in the injured glomeruli that incorporate more blunders in the annotation due to the identification difficulty from morphology changing.

\subsection{Ablation study}
The purpose of corrective learning is to alleviate the noise and distillate the correct information, so as to improve the model performance using lay annotation. Four designs of corrective learning with different utilization of similarity losses and confidence losses were evaluated with lay annotation in Table.~\ref{tab:differentdesigns}. Each score is used in either an exponential function or a linear function (Eq.~\eqref{weight_W_S}), when multiplying and calculating the loss function (Eq.~\eqref{loss}). The bold configuration was selected as the final design.

\begin{table}[h]
\caption{Ablation study on different molecular-oriented corrective learning design.}
%\centering
\begin{center}
\begin{adjustbox}{width=0.7\textwidth}
%\begin{tabular}{p{23mm}p{23mm}|p{18mm}p{18mm}p{18mm}}
\begin{tabular}{p{23mm}p{23mm}|ccc}
\hline
Confidence score &  Similarity score & Podocyte F1 & Masengial F1 & Average F1\\
\hline
    Linear & Linear & 0.7255    & 0.6843    & 0.7049 \\
    Linear & Exponent & 0.7300    & 0.6987    & 0.7144 \\
    Exponent & Linear & \textbf{0.7411}    & \textbf{0.7028}    & \textbf{0.7219} \\
    Exponent & Exponent & 0.7304  & 0.6911    & 0.7108 \\
\hline
\end{tabular}
\end{adjustbox}
\end{center}
\label{tab:differentdesigns} 
\end{table}

\section{Conclusion}
In this work, we proposed a holistic, molecular-empowered learning solution to alleviate the difficulties of developing a multi-class cell segmentation deep learning model from the expert level to the lay annotator level, enhancing the accuracy and efficiency of cell-level annotation. An efficient corrective learning strategy is proposed to offset the impact of noisy label learning from lay annotation. The results demonstrate the feasibility of democratizing the deployment of a pathology AI model while only relying on lay annotators.

% ~\\
\noindent\textbf{Acknowledgements}. 
This work is supported in part by NIH R01DK135597(Huo), DoD HT9425-23-1-0003(HCY), and NIH NIDDK DK56942(ABF).
%\section{Acknowledgements}
% *****************

% *****************  \\
% Dr. Wheless is funded by grants from the Skin Cancer Foundation and the Dermatology Foundation.  The results published here are in whole or part based upon data generated by the TCGA Research Network: https://www.cancer.gov/tcga.

%The proposed method can be potentially applied to other circle/ball shaped biomedical objects, such. as certain types of cells, organs, and nuclei.

% \textbf{Acknowledgements}:
% *****

% \clearpage
\bibliographystyle{splncs04}
\bibliography{main}




% LNCS Homepage, \url{http://www.springer.com/lncs}. Last accessed 4
% \end{thebibliography}
\end{document}